\documentclass[aps,pra,superscriptaddress,reprint,floatfix,notitlepage,balancelastpage,twocolumn,showpacs,reprint]{revtex4}
\usepackage{graphicx,amsmath,amssymb,bm,color,engord}
\usepackage[colorlinks,urlcolor=red,citecolor=red]{hyperref}
\usepackage[caption=false]{subfig}

\newcommand{\nf}{n_{\rm F}}
\newcommand{\as}{a_s}

\newcommand{\kh}{\hat{k}}
\newcommand{\qh}{\hat{q}}

\newcommand{\tc}{{T_{\rm c}}}

\newcommand{\ef}{\epsilon_{\rm F}}
\newcommand{\kf}{k_{\rm F}}

\newcommand{\curO}{{\cal O}}

\newcommand{\kb}{k_{\rm B}}

\newcommand{\bk}{{\bf k}}
\newcommand{\bq}{{\bf q}}
\newcommand{\bQ}{{\bf Q}}

\newcommand{\be}{\begin{equation}}
\newcommand{\ee}{\end{equation}}
\newcommand{\bea}{\begin{eqnarray}}
\newcommand{\eea}{\end{eqnarray}}
\newcommand{\bse}{\begin{subequations}}
\newcommand{\ese}{\end{subequations}}

\begin{document}

\title{Induced  superfluidity  of imbalanced  Fermi gases near unitarity}
\author{Kelly R. Patton}
\email[]{kpatton@physast.uga.edu}\affiliation{Seoul National University, Department of Physics and Astronomy\\ Center of Theoretical Physics, 151-747 Seoul, Korea}

\author{Daniel E. Sheehy}
\email[]{sheehy@phys.lsu.edu}\affiliation{Department of Physics and Astronomy, Louisiana State University, Baton Rouge, Louisiana 70803 }
\date{March 28, 2012}
\begin{abstract} 
The induced intraspecies interactions among the majority species, mediated by
the minority species, is computed for a population-imbalanced two-component Fermi gas.  
Although the Feshbach-resonance mediated interspecies interaction is dominant
for equal populations, leading to singlet $s$-wave pairing, we find
that in the strongly imbalanced regime the induced intraspecies interaction
leads to $p$-wave pairing and superfluidity of the majority species.  Thus,
we predict that the observed spin-polaron Fermi liquid state in this regime
is unstable to $p$-wave superfluidity, in accordance with the results of Kohn and Luttinger,
below a temperature that, near unitarity, we find to be within current experimental
capabilities.    Possible experimental signatures of the
$p$-wave state using radio-frequency spectroscopy as well as density-density 
correlations after free expansion are presented.  
\end{abstract}
\pacs{05.30.Fk, 03.75.Ss, 67.85.-d, 32.30.Bv}
\maketitle

\section{Introduction} The extraordinarity variety of tunable \lq\lq knobs\rq\rq\
in cold atomic gas experiments has yielded a wide assortment of correlated phases of
matter ranging from superfluid and Mott insulating phases of bosonic atoms, to 
superfluid and nonsuperfluid phases of fermionic atoms~\cite{KetterleZwierlein,GurarieAnnalsPhys07,BlochRMP08,GiorginiRMP08}.
In the fermionic case the experimentally 
available knobs include the interactions
among two species of fermion, their relative 
densities~\cite{ZwierleinScience06,PartridgeScience06,Shin2006,Partridge2006prl}, and also the effective 
spatial dimension, which can be controlled via an appropriate confining trapping potential.  

Our present focus is on the case of a three-dimensional two-species
Fermi gas (labeled by the spin index $\sigma = \uparrow,\downarrow$) interacting via a magnetic field-tuned Feshbach resonance. 
In the balanced case of equal densities of the two species, and as a function of
the Feshbach resonance detuning that essentially controls $\frac{1}{a_s}$ with 
$a_s$ the $s$-wave scattering length, 
the atomic gas undergoes the well-known crossover from a Bose-Einstein condensate (BEC) of
tightly bound molecular pairs
at $\frac{1}{\kf a_s}\agt 1$ ,
 through the strongly correlated unitary regime at $\frac{1}{\kf a_s} \simeq 0$, to a weakly coupled
Bardeen-Cooper-Schrieffer (BCS) superconductor of Cooper pairs at $\frac{1}{\kf a_s}\alt-1$.  Here, $\kf \propto n^{1/3}$
is the Fermi wavevector (with $n$ the total density), the inverse of which characterizes the typical interparticle spacing.  Importantly, the low-temperature
state of this balanced Fermi gas is believed to undergo no symmetry-changing phase transitions
as the BEC-BCS crossover is traversed.

\begin{figure}
\includegraphics[width= 0.99\columnwidth]{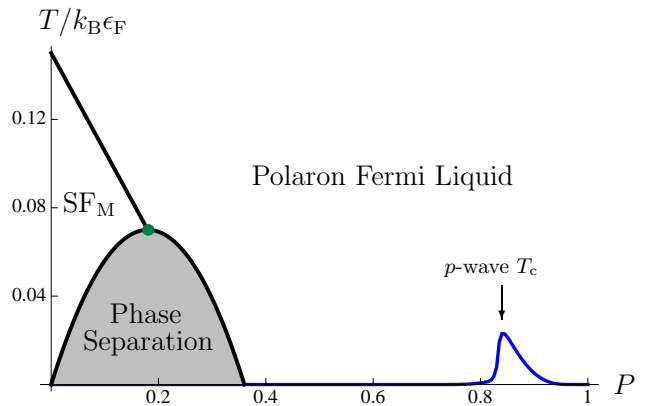}
\caption{(Color online) Proposed phase diagram of population imbalanced Fermi gases,
at unitarity $\frac{1}{\kf \as} = 0$,  as a function of temperature
$T$ (normalized to the Fermi energy $\ef$ multiplied by the Boltzmann constant $\kb$) and population imbalance
$P$,  showing regions of imbalanced  superfluid SF$_{\rm M}$, phase
separation (at temperatures below the tricritical point, a solid 
dot (green online)~\cite{Gubbels,Parish2007,SheehyTri}) polaron Fermi liquid, and a $p$-wave paired state.  The
principal question is whether the polaron Fermi liquid phase, in which
both spins-$\uparrow$ and spins-$\downarrow$ possess Fermi surfaces
(despite the strong interactions), persist to $T\to 0$.  The black phase
boundaries are only sketched but consistent with the experimentally
determined phase diagram of Ref.~\onlinecite{ShinNature}.  Below 
the blue curve, derived here,  we find
an instability towards $p$-wave pairing of the spin-$\uparrow$ Fermi
surface.  }\label{phasefig}
\end{figure}

The behavior of strongly interacting 3D Fermi gases in the {\it imbalanced\/} case is
considerably richer, with numerous phases having been predicted~\cite{Bedaque,Son,Pao,SR2006,Chien,Parish2007,SR2007,SR2007Comment} to occur
(as recently reviewed in Refs.~\onlinecite{RS2010,ChevyMora2010}),
 including the long sought-after  Fulde-Ferrell-Larkin-Ovchinnikov (FFLO) state~\cite{FF,LO}, in which 
the superfluid spontaneously develops a spatial modulation in the local pairing amplitude to 
accomodate the excess fermions of the majority species~\cite{Yoshida,LeoAshvin,LeoPRA}.  Unfortunately, the
FFLO phase has not yet been observed in 3D imbalanced Fermi gases (although evidence for a 1D
analogue of this state has been found~\cite{LiaoRittner}).  Instead, under an imposed density imbalance, characterized by 
the population imbalance $P = \frac{n_\uparrow - n_\downarrow}{n_\uparrow + n_\downarrow}$ (note we always
assume the densities $n_\sigma$ of species $\sigma$ satisfy $n_\uparrow\geq n_\downarrow$),
interacting Fermi gases enter a regime of phase separation at moderate $P$ followed by an imbalanced
nonsuperfluid phase at large $P$, as seen in the phase diagram Fig.~\ref{phasefig}.

In the strongly imbalanced limit ($P\to 1^-$) our system amounts to considering the phase resulting from adding
a few spins-$\downarrow$ to a Fermi sea of the spins-$\uparrow$.  In this
strongly imbalanced regime, within the simplest picture there are two possible fates of such an added spin down: 
It can either  form a molecular bound state with one spin-$\uparrow$, or it can remain unpaired. 
 In the former case, occurring at  $\as >0$ but $\kf \as \alt 1$
(i.e.~the BEC limit), as more spins-$\downarrow$ are added more molecules will develop and, presumably, condense.
Our interest is in the latter
regime, in which (within this simple picture) no interspecies  pairing occurs and condensation is suppressed.

From one point of view, the inability to establish pairing in this regime is simply traced to the fact that, with an imposed
density imbalance, not all the majority spins-$\uparrow$ have a   minority spin-$\downarrow$
fermion to pair with.  More precisely, the presence of a density imbalance
implies a concomitant Fermi surface (and Fermi energy) mismatch, so that the formation of low-energy Fermi surface 
pairing correlations is interrupted, leaving an imbalanced interacting Fermi liquid state~\cite{lobo2006,PunkPRL07,VeillettePRA08}
that, apparently, lacks any superfluid order even at low temperatures.

How do the strong attractive interspecies interactions manifest themselves, given the inability of this system to
form $s$-wave Cooper pairs and condense?  Recent experiments show evidence of the formation of Fermi polarons,
in which a cloud of the spins-$\uparrow$ form around each spin-$\downarrow$, leading to an observable
shift in the spin-$\downarrow$ chemical potential~\cite{SchirotzekPRL09,Nascimbene} in agreement with 
theoretical predictions~\cite{ChevyPRA06,CombescotPRL07,CombescotPRL08}.  The central question studied here 
concerns whether, at $T\to 0$, this polaron phase of matter persists or whether another
broken-symmetry phase emerges. 

One motivation for this possibility is the well-known results of Kohn and Luttinger~\cite{KohnPRL65, LuttingerPR66},
who showed that interacting Fermi liquid phases are generally unstable to pairing in some angular momentum
channel at low $T$.  Given that the polaron state of imbalanced Fermi gases is, at its heart, 
essentially a Fermi liquid for both species (exhibiting, for example, a sharp Migdal discontinuity in
the momentum occupation at the Fermi surface), we generally expect the Kohn-Luttinger mechanism to 
hold, yielding odd angular momentum  {\it intraspecies\/} 
Cooper pairing at both the spin-$\uparrow$ and spin-$\downarrow$
Fermi surfaces for $T\to 0$~\cite{BulgacPwavePRL,Patton}.   The odd-angular momentum pairing requirement
follows from the  Pauli exclusion principle; while the simplest such state has $p$-wave symmetry (assumed
here), more generally any odd $\ell$ is possible.  

In fact, $p$-wave pairing is also possible in the deep BEC limit of imbalanced gases via  a somewhat 
different mechanism~\cite{BulgacPwavePRL,BulgacPRA09}.
As noted above, in the deep BEC limit tightly bound molecular pairs form even at very large imbalance 
(only vanishing when $n_\downarrow \to 0$).  If we imagine decreasing $P$ from unity in this limit,
within a mean-field picture two possibilities emerge~\cite{SR2006,Parish2007,SR2007}: Firstly, the system 
can form a homogeneous polarized magnetic superfluid phase (SF$_{\rm M}$ in Fig.~\ref{phasefig}), in which
such molecular pairs coexist with a Fermi sea of the excess spins-$\uparrow$.  This possibility is found
in the  very deep BEC limit (i.e., $\frac{1}{\kf \as} >2.37$ within mean-field theory, with the tricritical point
of Fig.~\ref{phasefig} becoming a quantum tricritical point~\cite{Parish2007} at $\frac{1}{\kf \as} =2.37$).
Secondly, closer to unitarity, the system can form a phase separated mixture of  SF$_{\rm M}$ and imbalanced
normal phase.  Our point here is that, in the  SF$_{\rm M}$ phase, $p$-wave pairing can be induced among
the excess spins-$\uparrow$ by the molecular bosons, a mechanism studied in Refs.~\cite{BulgacPwavePRL,BulgacPRA09}.
Here, we work away from the regimes of phase separation and SF$_{\rm M}$, focusing on induced pairing
of the spins-$\uparrow$ mediated 
purely by the presence of the spins-$\downarrow$ (i.e., without any molecular pairing);
however, the relationship between these two regimes is an interesting problem for future research. 

In this paper we present the details of our calculation of the transition temperature below which 
 $p$-wave pairing is expected to occur within this mechanism, expanding upon our recent Rapid Communication~\cite{Patton}.  
This calculation can be summarized by the phase diagram
Fig.~\ref{phasefig} for imbalanced Fermi gases at unitarity, showing regions of polaron Fermi liquid,
imbalanced or \lq\lq magnetic\rq\rq\ superfluid  SF$_{\rm M}$, and phase separation.  Here, the
first-order  phase boundary enclosing the regime of phase separation and the second-order phase boundary 
separating the SF$_{\rm M}$ and polaron Fermi liquid phases are only sketched (i.e. are not the result
of a calculation).  However, we drew these phase boundaries to be consistent with the experimental 
results of Ref.~\onlinecite{ShinNature} to illustrate the fact that our maximum predicted transition
temperature for $p$-wave pairing (solid curve, blue online) is not much smaller than the temperature
scales characterizing these phase boundaries, suggesting that it may be possible to observe $p$-wave
pairing of strongly imbalanced Feshbach-resonant Fermi gases.  

 We remark that other recent work has
considered the effect of intraspecies interactions in imbalanced Fermi gases~\cite{Liao}; however, this work
assumed an intrinsic intraspecies interaction (i.e., in the Hamiltonian), while our work assumes
a vanishing {\it bare} intraspecies interactions.  The present problem of induced intraspecies interactions in 
imbalanced gases has previously been studied by Bulgac and collaborators~\cite{BulgacPwavePRL},
and by Nishida, the latter in the two-dimensional limit~\cite{NishidaAnalPhys09}.
  Additional recent work has studied the problem of
polaron-polaron interactions in imbalanced Fermi gases~\cite{Giraud}, although this work did not address the 
possibility of $p$-wave pairing.



%
%
%
%

This paper is organized as follows.
In Sec.~\ref{SEC:modelham} we start from the standard one-channel model Hamiltonian for
imbalanced Fermi gases and develop a general Green's-function formalism for studying 
intraspecies pairing in this setting.  In this section we show that intraspecies pairing
among the spins-$\uparrow$
is reflected by an off-diagonal component of an associated Nambu self energy that, in
turn, depends on a matrix vertex function that we proceed to approximate in the subsequent
sections. In Sec.~\ref{Perturbative results}, we approximate this vertex function to
quadratic order in perturbation theory; our results in this section are consistent with the 
original calculations of Kohn and Luttinger~\cite{KohnPRL65,LuttingerPR66} (who studied the 
general problem of induced interactions in fermions) and, more recently, Bulgac et al.~\cite{BulgacPwavePRL} 
(who studied induced interactions  in the present setting of imbalanced Fermi gases).  Importantly, the perturbative
formula for the transition temperature is invalid in the unitary regime where cold-atom experiments
focus; in the regime where it applies, this transition temperature is orders of magnitude too small
to be observable.   In Sec.~\ref{strongly interacting effective interaction}, we attempt to 
go beyond the perturbative limit to estimate the transition temperature for $p$-wave
pairing among the majority species in the unitary regime.   The classes of diagrams we keep for
the vertex function include both ladder and crossed ladder diagrams (which, as we show, recover
the conventional Fermi liquid theory of imbalanced gases), but also diagrams containing both
ladder and crossed-ladder subdiagrams (the latter required to have a nonzero transition temperature).  
In Sec.~\ref{sec:ltpg}, we use our results from Sec.~\ref{strongly interacting effective interaction}
to estimate the magnitude of the pairing gap at the spin-$\uparrow$ Fermi surface at $T\to 0$.  In Sec.~\ref{sec:ed},
we turn to the question of how the $p$-wave phase of imbalanced Fermi gases could be observed in
radio-frequency spectroscopy and density  correlation experiments before concluding in Sec.~\ref{sec:conc}.


\section{Model Hamiltonian and formalism}
\label{SEC:modelham}
Our aim is to derive the effective interactions  among one species of fermion, 
mediated by the other species of fermion, in a strongly imbalanced Fermi gas.
In the present section we begin by developing a Green's function formalism to
address this problem.
 Our starting point is the following one-channel
model Hamiltonian for a gas of two species of fermions (labeled by
$\sigma = \uparrow,\downarrow$)
interacting via an $s$-wave Feshbach resonance~\cite{GurarieAnnalsPhys07}
(note we take $\hbar = 1$):
%
%
%
%
%
\begin{align}
\label{Hamiltonian}
H&=\sum_{\sigma}\int d^{3}r\, \Psi^{\dagger}_{\sigma}({\bf r})\left[-\frac{\nabla^{2}}{2m}-\mu_{\sigma}\right]
\Psi^{}_{\sigma}({\bf r})\nonumber\\&+{\lambda}\int d^{3}r\,  \hat{n}_{\uparrow}({\bf r})\hat{n}_{\downarrow}({\bf r}),
\end{align}
 where the density $\hat{n}_{\sigma}({\bf r})  = \Psi^{\dagger}_{\sigma}({\bf r})\Psi^{}_{\sigma}({\bf r})$
and $\lambda$ is the strength of a short-ranged
pseudo-potential (approximated by a delta-function in real space). 
Here, $\mu_\sigma$ is the chemical potential of species $\sigma$; the density imbalance
can be considered to arise from an imbalance in the chemical potentials $\mu_\uparrow \neq \mu_\downarrow$.

The Hamiltonian Eq.~(\ref{Hamiltonian}) must be defined
along with a cutoff reflecting the short-distance properties
of the real physical interaction, with a corresponding scale  $d$.  Equivalently, 
this problem has a large momentum cutoff $\Lambda\approx 2\pi/d$ that
regularizes any divergent ultraviolet (UV) behavior coming from the
singular nature of the delta-like  pseudo-potential. 
In practice, as is well known~\cite{GurarieAnnalsPhys07}, this can be handled by exchanging the bare coupling $\lambda$
for the vacuum scattering length $\as$ that are related via 
\be
\label{lambda vs a}
\frac{1}{\lambda}=\frac{m}{4\pi \as}-\sum_{\bf k}^{\Lambda}\frac{1}{2\epsilon_{\bf k}}.
\ee
In the weak-coupling BCS limit $\lambda \to 0-$, this equation is solved by $\lambda = \frac{4\pi \as}{m}$, i.e,
we can neglect the final term on the right side of Eq.~(\ref{lambda vs a}).   In the unitary regime where $\as$ becomes
large, we use a different procedure: 
If we assume physical observables are independent of $\Lambda$, then to study 
 systems at fixed values of
$\as$ (which is experimentally controllable) it is valid to replace $\lambda$ by 
$\as$ [using Eq.~(\ref{lambda vs a})] in approximate
theoretical expressions and then take the limit $\Lambda \to \infty$.  This strategy, which is equivalent
to taking $\lambda \to 0-$ and $\Lambda \to \infty$ while holding $\as$ fixed via Eq.~(\ref{lambda vs a}), will naturally lead to our
inclusion of certain classes of Feynman diagrams.


As we have discussed,
the Kohn-Luttinger result implies that both 
the spin-$\uparrow$ and spin-$\downarrow$ Fermi surfaces of an imbalanced
Fermi gas are unstable 
at $T\to 0$.  This low-$T$ regime, possibly consisting of
two interpenetrating $p$-wave superfluids, is beyond the scope of this manuscript
and we leave it for future work.  Instead, we assume that, owing
to the large population imbalance, the transition
temperature for any $p$-wave pairing of the minority spins-$\downarrow$ is 
much smaller than the corresponding temperature for the majority spins-$\uparrow$,
as previously found in the weak-coupling BCS limit~\cite{BulgacPwavePRL}.

%

%

\begin{figure}
\includegraphics[width= 0.75\columnwidth]{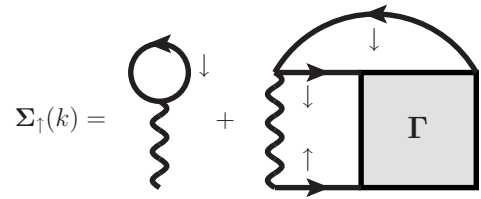}
\caption{\label{fig1}The formally exact diagrammatic matrix self-energy, Eq.\ \eqref{exact self-energy}.  
The spin-$\uparrow$ Green's function and vertex function $\boldsymbol{\Gamma}$ are Nambu matrices, 
while the 
spin-$\downarrow$ are normal scalar functions.  The interaction lines are
$\lambda\boldsymbol{\sigma}_{z}$, 
where $\boldsymbol{\sigma}_{z}$ is the third Pauli matrix.}
\end{figure}

To derive the effective induced interactions among the spins-$\uparrow$, mediated by
the spins-$\downarrow$, we derive self-consistent equations for the corresponding
Green's functions for the two species of fermions.  As discussed above, we
assume the spins-$\downarrow$ to be in an unpaired Fermi-liquid state, while
the spins-$\uparrow$ may possess pairing correlations.  For the latter,
accounting for such pairing correlations is best done by organizing
the spin-$\uparrow$ Green's functions using the Nambu notation:
\begin{equation}
\Phi^{}_{\uparrow}({\bf r})=\left(\begin{array}{c}\Psi_{\uparrow}({\bf r}) 
\\\Psi^{\dagger}_{\uparrow}({\bf r})\end{array}\right).
\end{equation}
 We can then define the imaginary time ordered  Nambu matrix Green's function  for spins-$\uparrow$
 \begin{equation}
 {\bf G}_{\uparrow}({\bf r},\tau)=-\langle T\Phi^{}_{\uparrow}({\bf r},\tau)\Phi^{\dagger}_{\uparrow}(0,0)\rangle^{}_{H},
 \end{equation}
with matrix elements 
 \begin{align}
 \label{Green's function} {\bf G}_{\uparrow}({\bf
r},\tau)&=\left(\begin{array}{cc}G_{\uparrow}({\bf r},\tau) &
F_{\uparrow}({\bf r},\tau) \\ F^{*}_{\uparrow}({\bf r},\tau) &
-G_{\uparrow}(-{\bf
r},-\tau) \end{array}\right).
 \end{align}
 The normal Green's functions are given by
 \begin{equation}
 G^{}_{\uparrow}({\bf r},\tau)=
-\langle T\Psi^{}_{\uparrow}({\bf r},\tau)\Psi^{\dagger}_{\uparrow}(0,0)\rangle^{}_{H},
 \end{equation}
 and the anomalous ones are
  \begin{align}
& F^{}_{\uparrow}({\bf r},\tau)=-\langle T\Psi^{}_{\uparrow}({\bf r},\tau)\Psi^{}_{\uparrow}(0,0)\rangle^{}_{H},
 \nonumber \\&F^{*}_{\uparrow}({\bf r},\tau)=-\langle T\Psi^{\dagger}_{\uparrow}({\bf r},\tau)\Psi^{\dagger}_{\uparrow}(0,0)\rangle^{}_{H}.
 \end{align}
For notational convenience, we further introduce the four-vector $k=({\bf k},i\omega_{n})$. In Fourier-Matsubara space the spin-$\uparrow$ Green's function satisfies a matrix analog of the Dyson equation
 \begin{equation}
 \label{matrix Dyson equation}
 {\bf G}_{\uparrow}(k)={\bf G}_{0,\uparrow}(k)+{\bf G}_{0,\uparrow}(k)\boldsymbol{\Sigma}_{\uparrow}(k){\bf G}_{\uparrow}(k),
 \end{equation}
 where 
 \begin{align}
 {\bf G}_{0,\uparrow}(k)&=\left(\begin{array}{cc}G_{0,\uparrow}(k) & 0 \\0 & -G_{0,\uparrow}(-k) \end{array}\right),
\nonumber\\&=\left(\begin{array}{cc}\frac{1}{i\omega_{n}-\xi_{{\bf k}\uparrow}} & 0 \\0 & \frac{1}{i\omega_{n}+\xi_{{\bf k}\uparrow}} \end{array}\right),
 \end{align}
 with $\xi_{{\bf k}\sigma}=\epsilon_{\bf k}-\mu_{\sigma}$, and 
 \begin{equation}
 \label{Nambu self-energy}
 \boldsymbol{\Sigma}_{\uparrow}(k)=\left(\begin{array}{cc}\Sigma_{\uparrow}(k) & \Delta_{\uparrow}(k) \\\Delta^{*}_{\uparrow}(k) & -\Sigma_{\uparrow}(-k)\end{array}\right),
 \end{equation}
 is the matrix self-energy. The appearance of nonzero off-diagonal terms in the self-energy indicates the presence of a superfluid state, with order parameter or gap function $\Delta_{\uparrow}(k)$. 
The self energy satisfies
 \begin{align}
 \label{exact self-energy}
 \boldsymbol{\Sigma}_{\uparrow}(k)&=\lambda \boldsymbol{\sigma}_{z}\sum_{q}G_{\downarrow}(q)+\lambda \boldsymbol{\sigma}_{z}\sum_{q,q'}{\bf G}_{\uparrow}(q)G_{\downarrow}(q')\times\nonumber\\&\times \boldsymbol{\Gamma}(q,q',q+q'-k,k)G_{\downarrow}(q+q'-k),
 \end{align}
where  $\boldsymbol{\sigma}_{z}$ is a
Pauli matrix, $\boldsymbol{\Gamma}$ is the reducible two-particle
matrix vertex function (apart from energy and momentum conserving
delta functions), and the summation is  $\sum_{q}\equiv (\beta{\sf
V})^{-1}\sum_{{\bf q}}\sum_{i\nu_{n}}$, with $\beta^{-1}=k_{\rm B}T$
and $\sf V$ the system volume (which, henceforth, we will set to unity).  Diagrammatically, Eq.~(\ref{exact self-energy}) is shown in Fig.~\ref{fig1}. 

The spin-down Green function 
$G^{}_{\downarrow}({\bf r},\tau)$ satisfies a similar set of equations (but is assumed to be unpaired); in terms of the 
spin-$\downarrow$ fermion operators it is defined as 
 \begin{equation}
G^{}_{\downarrow}({\bf r},\tau)=-\langle T\Psi^{}_{\downarrow}({\bf r},\tau)\Psi^{\dagger}_{\downarrow}(0,0)\rangle_{H}.
\end{equation}
Since the Greens's function depends on $\boldsymbol{\Sigma}_{\uparrow}(k)$ via the Dyson equation [Eq.~(\ref{matrix Dyson equation})],
this, along with Eq.~(\ref{exact self-energy}) amount to self-consistent equations for the self-energy.  For pairing to be stable, 
we must find a solution to these equations possessing a nonzero off-diagonal component $\Delta_{\uparrow}(k)$.  Formally, the only assumption we have made thus
far is  that the spins-$\downarrow$ are unpaired (i.e., they possess no off-diagonal component to their self energy); in practice to 
proceed we must make a physically-motivated approximation for the Bethe-Salpeter equation satisfied by  $\boldsymbol{\Gamma}$.

\section{Leading-order perturbation theory}
\label{Perturbative results}
Having set up the general formalism for computing the self-consistent pairing amplitude
for intraspecies pairing correlations among the spins-$\uparrow$, in the present section we 
take the weak-coupling perturbative approximation $\lambda \to 0-$ in which these equations
simplify.  This will allow us to easily identify the effective interaction among
the spins-$\uparrow$; as we show below, a similar structure will hold when we sum 
diagrams to all orders in $\lambda$ (to access the unitary regime).  Our results in this
section are consistent with those of Ref.~\onlinecite{BulgacPwavePRL}.

 In the weakly interacting BCS limit we can expand perturbatively in $\lambda$, or equivalently
in $\as = m\lambda/4\pi$.   In fact, to obtain a nonzero result for the intraspecies pairing we 
must keep terms of order $\lambda^2$.  The reason for this is that, to $\curO(\lambda)$, 
the self energy is simply given by the first term in Eq.~\eqref{exact self-energy}, the
Hartree contribution,  $\boldsymbol{\Sigma}_{\uparrow}(k)=\lambda\boldsymbol{\sigma}_{z}n_{\downarrow}$,
where $n_{\downarrow}$ is the density of spins-$\downarrow$.  At this level of approximation, 
the  Nambu self-energy is diagonal and no pairing is induced.

To obtain results valid to quadratic order, $\curO(\lambda^2)$, it is sufficient to approximate the 
vertex function by the bare interaction, 
\be
\boldsymbol{\Gamma}=-\lambda
\boldsymbol{\sigma}_{z} ,
\ee
where the presence of the Pauli matrix is due to the fact that we need to write the interaction in Nambu space.  
This gives for the self energy:
\begin{equation}
\label{second order self-energy}
 \boldsymbol{\Sigma}_{\uparrow}(k)=\lambda \boldsymbol{\sigma}_{z}n_{\downarrow}-\lambda^{2}\boldsymbol{\sigma}_{z}
 \sum_{q,q'}{\bf G}_{\uparrow}(q)\boldsymbol{\sigma}_{z}G_{\downarrow}(q')G_{\downarrow}(q+q'-k).
\end{equation}
Here and in the next section (when we sum diagrams to all orders in $\lambda$), we'll neglect the diagonal 
component of the self-energy (simply assuming it amounts to a chemical potential shift).  Within this simplifying approximation,
the Green's function for the spins-$\uparrow$ can be written as 
\begin{equation}
 \label{Green's function with no normal self-energy}
 {\bf G}_{\uparrow}({\bf k},i\omega_{n})=\frac{1}{(i\omega_{n})^{2}-E^{2}_{\bf k}}\left(\begin{array}{cc}i\omega_n+\xi_{{\bf k}\uparrow} & \Delta_{\uparrow}({\bf k},i\omega_n) \\\Delta^{*}_{\uparrow}({\bf k},i\omega_n) & i\omega_n-\xi_{{\bf k}\uparrow}\end{array}\right),
 \end{equation}
 where $E_{\bf k}(i\omega_{n})=\sqrt{\xi^{2}_{{\bf k}\uparrow} +\Delta^{*}_{\uparrow}({\bf k},i\omega_n) \Delta^{}_{\uparrow}({\bf k},i\omega_n)}$. Comparing  Eqs.\ \eqref{Green's function} 
and \eqref{Green's function with no normal self-energy} one can read off the relationship between the anomalous propagator and gap function 
 \begin{equation}
 \label{anomalous propagator}
 F_{\uparrow}({\bf k},i\omega_{n})=-\frac{\Delta_{\uparrow}({\bf k},i\omega_n)}{\omega^{2}_{n}+E^{2}_{\bf k}}. 
 \end{equation}

\begin{figure}\label{fig3}
\includegraphics[width= 0.6\columnwidth]{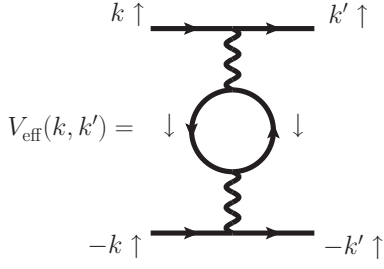}%
\caption{The induced interaction among the spins-$\uparrow$, mediated by the spins-$\downarrow$, in the weak-coupling 
 BCS limit. The internal loop corresponds to density fluctuations of the minority spin.   The external legs are only present to label the incoming and outgoing energy and momentum. }
\end{figure}

This approximation also leads to a simple form for the upper-right off-diagonal component of 
Eq.~\eqref{second order self-energy}:
\bea
\label{general gap equation}
&&\hspace{-1cm}\Delta_{\uparrow}({\bf k},i\omega_{n})=\frac{-1}{\beta}\sum_{{\bf q},i\nu_{n}}
V^{}_{\rm eff}({\bf k},{\bf q},i\omega_{n},i\nu_{n}) \frac{\Delta_{\uparrow}({\bf q},i\nu_{n})}{\nu^{2}_{n}+E^{2}_{\bf q}},
\\
\label{perturbative interaction}
&&V_{\rm eff}( k,q)\equiv\lambda^{2}\sum_{p}G_{\downarrow}({k}-q+{ p})G_{\downarrow}({p}),
\eea
where, in the second line, we used a short hand notation 
$V_{\rm eff}( k,q) \equiv V^{}_{\rm eff}({\bf k},{\bf q},i\omega_{n},i\nu_{n})$.
Thus, the similarity of the resulting expression to a standard gap equation for pairing 
has allowed us to identify an effective interaction $V_{\rm eff}( k,k')$ that is plotted
diagramatically in Fig.~\ref{fig3}.

We proceed by 
approximating the full spin-$\downarrow$ Green's function in \eqref{perturbative interaction} by noninteracting ones, i.e., 
$G_{\downarrow}({\bf k},i\omega_{n})\rightarrow G_{0,\downarrow}({\bf k},i\omega_{n})=(i\omega_{n}-\xi_{{\bf k}\downarrow})^{-1}$, 
neglecting the frequency 
dependence of $V_{\rm eff}( k,k')$ (setting the Matsubara frequencies to zero), and assuming that $\Delta_{\uparrow}({\bf k},i\omega_{n}) = \Delta_{\uparrow}({\bf k})$, i.e., 
it is indepenent of frequency.  

\begin{figure}
\includegraphics[width= \columnwidth]{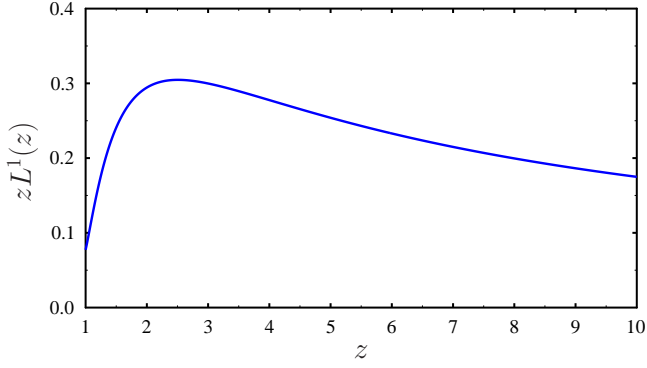}%
\caption{(Color online) The $p$-wave projection of the Lindhard Function, see Eq.\ \eqref{real perturbative  interaction}, as a function of density imbalance $z=k_{{\rm F}\uparrow}/k_{{\rm F}\downarrow}$.  The peak near $z\approx2$ occurs as a result as the turning off of particle-hole excitations with a transferred momentum  larger than the diameter of the spin-$\downarrow$ Fermi surface, while at smaller wave vectors the Lindhard function is approximately flat, leading to a vanishingly small $p$-wave projection.}\label{lind} 
\end{figure}

Evaluating the remaining Matsubara sum in Eq.~(\ref{general gap equation}), 
we arrive at the following gap equation that is of the standard form:
\begin{equation}
 \Delta_{\uparrow}(\bk)=-\sum_{{\bf q}}V_{\rm
eff}(\bk,\bq) \frac{\tanh\big(\beta E^{}_{\bf q}/2\big)}{2E^{}_{\bf
q}}\Delta_{\uparrow}({\bf q}),
\label{on-shell gap equation}
\end{equation}
where now the induced interactions (mediated by density fluctuations of the spins-$\downarrow$)
are proportional to the Lindhard function:
\bea
\label{real perturbative  interaction}
V_{\rm eff}(\bk,\bq)&=&-\Big(\frac{4\pi \as}{m}\Big)^{2}N_{\downarrow}(\epsilon_{{\rm F}\downarrow}) L(|\bk - \bq|/2k_{{\rm F}\downarrow})),
\\
L(x)&=&\frac{1}{2}+\frac{1-x^{2}}{4x}\ln\left|\frac{1+x}{1-x}\right|,
\eea
where we have replaced $\lambda \to 4\pi \as/m$ as noted above (valid in the weak-coupling BCS regime).
Here, $k_{{\rm F}\sigma}$ is the Fermi wavevector 
for species $\sigma$, satisfying $n_\sigma = \frac{k_{{\rm F} \sigma}^3}{3\pi^2}$,  and 
\be
\label{nsigma}
N_{\sigma}(\epsilon_{{\rm F}\sigma})=mk_{{\rm F}\sigma}/(2\pi^{2}),
\ee 
is the  density of states at the spin-$\sigma$ Fermi surface.

We now decompose the gap potential and effective
interaction into angular momentum channels using 
\bea\label{angular momentum channels}
V_{\rm eff}({\bf k},{\bf k}')&=&\sum_{l=0}^{\infty}(2l+1)v^{l}_{k,k'}P_{l}(\hat{\bf k}\cdot\hat{\bf k}') ,
\\
\Delta_{\uparrow}({\bf k})&=&\sum_{l=0}^{\infty}(2l+1)\Delta^{l}_{k\uparrow}P_{l}(\hat{\bf z}\cdot\hat{\bf k}),
\eea
where $P_{l}(x)$ are the Legendre polynomials.  The transition temperature for each angular momentum channel is 
then determined by the solution to 
\begin{equation}
\label{gap equation}
\Delta^{l}_{k\uparrow}=-\sum_{{\bf q}}v^{l}_{k,q} 
\frac{\tanh\big(\beta \xi_{{\bf q}\uparrow}/2\big)}{2\xi_{{\bf q}\uparrow}}\Delta^{l}_{q\uparrow},
\end{equation}
where we have assumed a continuous transition where the gap potential vanishes; in this limit
we replace $E_{\bf q}\rightarrow \xi_{{\bf q}\uparrow}$.  To proceed, we note that we are
interested in the onset of $p$-wave ($l =1$) pairing at the spin-$\uparrow$ Fermi surface.  Thus, we set $k\to 
k_{{\rm F}\uparrow}$ in Eq.~(\ref{gap equation}), and henceforth choose $l =1$.  The summation over $\bq$ 
is then dominated by the region where $q \simeq k_{{\rm F}\uparrow}$. 
Using that the function $v^{1}_{k,q}$ is only nonzero for $k$ and $q$ within $~k_{{\rm F}\downarrow}$ of each
other, and converting the sum to an integral (introducing the density of states), we obtain
\begin{equation}
\Delta^{1}_{\uparrow}(\epsilon_{{\rm F}\uparrow})\approx-v^{1}_{k_{{\rm F}\uparrow},k_{{\rm F}\uparrow}} 
N_{\uparrow}(\epsilon_{{\rm F}\uparrow})\int\limits_{-\epsilon_{{\rm F}\downarrow}}^{\epsilon_{{\rm F}\downarrow}} 
d\epsilon\, \frac{\tanh\big(\beta \epsilon/2\big)}{2\epsilon}\Delta^{1}_{\uparrow}(\epsilon),
\end{equation}
for the pairing gap near the assumed continuous transition.  The corresponding transition temperature 
is:
\begin{equation}
\label{transition temperature}
k_{\rm B}T_{{\rm c}}\approx 
\frac{2\epsilon_{{\rm F}\downarrow}e^{\gamma}}{\pi}\exp\Big[\frac{1}{v^{1}_{k_{{\rm F}\uparrow},k_{{\rm F}\uparrow}} 
N_{\uparrow}(\epsilon_{{\rm F}\uparrow})}\Big],
\end{equation}
where $\gamma=0.577\ldots$ is the Euler-Mascheroni constant, giving our result for the 
transition temperature expressed in terms of the $\ell = 1$ projection
of the induced interactions in the weak-coupling regime $\as\to 0^-$.

The final step in 
the perturbative analysis is to obtain the $p$-wave
component of the perturbative effective interaction, Eq.~(\ref{real perturbative  
interaction}). Then, we have for the dimensionless 
effective induced interaction [appearing in the argument of the exponential function of 
Eq.~(\ref{transition temperature})] \cite{BulgacPwavePRL}:
\begin{equation}
\label{perturbative p-wave interaction}
v^{1}_{k_{{\rm F}\uparrow},k_{{\rm F}\uparrow}} N_{\uparrow}(\epsilon_{{\rm F}\uparrow})=-\frac{4(k_{{\rm F}\downarrow}\as)^{2}}{\pi^{2}}zL^{1}(z),
\end{equation}
where $z=k_{{\rm F}\uparrow}/k_{{\rm F}\downarrow}$ and 
\begin{equation}
L^{1}(z)=\frac{5z^{2}-2}{15z^{4}}\ln\left|1-z^{2}\right|-\frac{z^{2}+5}{30z}\ln\left|\frac{1-z}{1+z}\right|-\frac{z^{2}+2}{15z^{2}},
\end{equation}
 is the $p$-wave projection of the Lindhard function
\begin{equation}
\label{perturbative p-wave}
L^{1}(z)=\int\limits_{0}^{\pi}d\theta\, \cos\theta L(|\hat{{\bf k}}-\hat{{\bf k}'}|z/2),
\end{equation} with $\hat{\bf k}\cdot\hat{\bf k}'=\cos(\theta)$. 
The perturbative formula for $T_{\rm c}$, given by inserting Eq.~(\ref{perturbative p-wave interaction}) 
into Eq.~(\ref{transition temperature}) thus yields a result that vanishes exponentially as $k_{\rm B}T_{{\rm c}}~  \exp[-c/(k_{\rm F}\as)^2]$
with $c>0$; if this estimate is correct, then such $p$-wave pairing is probably not experimentally observable in the weak-coupling
perturbative regime. 

 For fixed $k_{{\rm F}\downarrow}\as$, the density imbalance dependence
enters through $zL^{1}(z)$, that we plot in Fig.~\ref{lind}.  We see that this quantity
shows a maximum value near $z\approx2$, leading to
a peak in the $p$-wave transition temperature  near a polarization of
$P\approx 0.77$; we find a similar peak in $\tc$ in our unitary-regime results to follow.

In the next section we proceed to derive a formula for  $T_{\rm c}$ which goes beyond the weak-coupling regime.  Our result 
 is of the same form as Eq.~(\ref{transition temperature})  but with a more complicated expression for the effective induced interactions
$v^{1}_{k_{{\rm F}\uparrow}k_{{\rm F}\uparrow}}$, including contributions from all orders of $\lambda$.   Our final result for $T_{\rm c}$
in fact reduces to the perturbative result of this section when we take the weak-coupling limit $\as\to 0- $; however, in the regime
where they agree (far to the right of the displayed area of Fig.~\ref{fig10}), $T_{\rm c}$ is many orders of magnitude smaller than the 
maximum transition temperature shown in Fig.~\ref{phasefig}.

\section{Beyond leading order}
\label{strongly interacting effective interaction}

In the preceding section, we showed that,  within leading-order perturbation theory, there is a 
transition to a $p$-wave superfluid of the majority species of a population-imbalanced Fermi gas, although the predicted
perturbative temperature is vanishingly small.  In the present section, we show how additional classes of diagrams, occurring 
near the unitary regime, can lead to an enhanced $T_{\rm c}$.  
To investigate the induced interaction in the strongly interacting regime, near unitarity, one must sum an infinite number of diagrams, 
to all orders of the bare interaction $\lambda$, contributing to the Nambu self-energy Eq.~\eqref{exact self-energy}.  We begin with the 
conventional  ${\sf T}$-matrix approximation, which, when extended to a system with pairing correlations, includes ladder and crossed-ladder diagrams.

\begin{figure}
\includegraphics[width= \columnwidth]{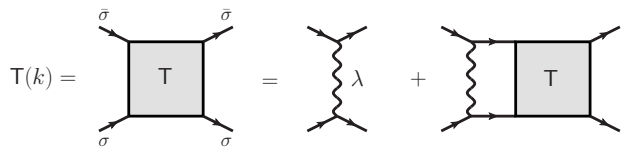}%
\caption{ The ${\sf T}$-matrix, which describes the ladder sum of two
particles interacting through the bare potential infinitely many
times. Here, the solid lines are Greens functions and the wavy line is the interaction $\lambda$.}\label{tee}
\end{figure}

\subsection{Ladder plus crossed ladder approximation} 
\begin{figure}
\includegraphics[width= \columnwidth]{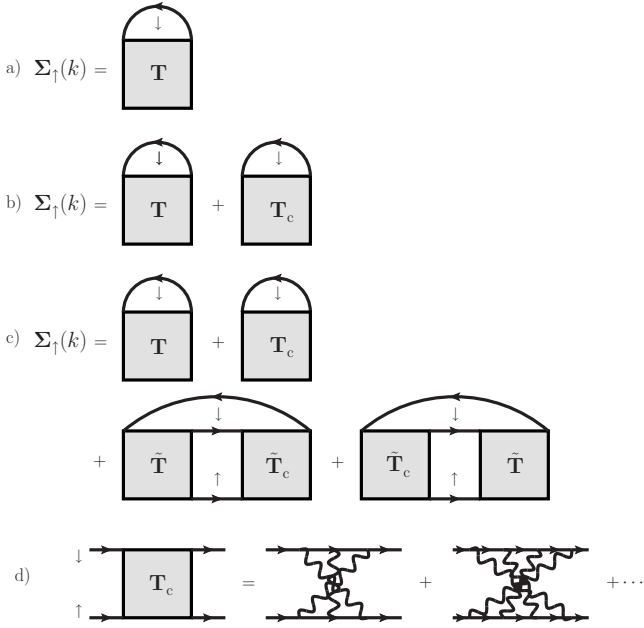}%
\caption{The diagrammatic contributions of the spin-$\uparrow$ Nambu
self-energy at various levels of approximations. Where $\bf T$,
defined by Eq.~\eqref{matrix T-matrix}, corresponds to the Nambu
matrix generalization of the $\sf T$-matrix, shown in Fig.~\ref{tee},
where all spin-$\uparrow$ Green's functions are in Nambu space while
the spins-$\downarrow$ are scalar functions and interaction lines
correspond to $\lambda\boldsymbol{\sigma}_{z}$. Here ${\bf T}_{\rm
c}$, given by Eq.~\eqref{mc t matrix}, is the Nambu generalized,  sum
of the so-called maximally crossed diagrams, shown in
panel d,  $\tilde{\bf T}=\boldsymbol{\Pi}_{A}{\bf T}$
and $\tilde{\bf T}_{\rm c}=-\boldsymbol{\Pi}^{-1}_{B}{\bf T}_{\rm c}$,
with $\boldsymbol{\Pi}_{A}$ and $\boldsymbol{\Pi}^{}_{B}$  defined
by Eq.~\eqref{Pi A} and Eq.\eqref{Pi B} respectively. }\label{fig6}
\end{figure}

Recent work has found that
the nonsuperfluid Fermi liquid phase of imbalanced Fermi gases is well
described by the so-called ${\sf T}$-matrix approximation for the
self-energy or more specifically for the vertex
\cite{PunkPRL07,CombescotPRL08,VeillettePRA08}, depicted diagrammatically in 
Fig.~\ref{tee}.  In the case of unpaired spins-$\uparrow$, 
this approximation amounts to summing the repeated  interaction of a spin-$\uparrow$
particle with a single spin-$\downarrow$ particle-hole bubble, as shown diagramatically in 
the self energy shown in Fig.~\ref{fig6}(a).

%
%

Our aim is to generalize the  $\sf{T}$-matrix approximation to imbalanced Fermi gases by including the possibility 
of pairing for the spins-$\uparrow$.  The simplest way to do this is to keep the same series of diagrams, 
but replace the normal Green's function $G_\uparrow(k)$ with the corresponding Nambu Green's function,
 i.e., considering the same ladder series but with $G_\uparrow(k) \to {\bf G}_\uparrow(k)$.  
%
%
 We refer to this Nambu $\sf{T}$-matrix with the bold symbol ${\bf T}(k)$. 
Formally it is the same as Fig.~\ref{tee} but with the spin-$\uparrow$ Green's functions possessing
Nambu structure (while the spin-$\downarrow$ Green's function is still a normal Green's function);
 additionally the coupling (wavy line)  is given by $\lambda \boldsymbol{\sigma}_z$.  The corresponding
self energy is given in  Fig.\ref{fig6}(a).

    However, this ladder series is not sufficient for our purposes, as can be seen by noting that it 
does not even reproduce the standard  $\sf{T}$-matrix approximation
in the limit $\Delta_\uparrow(k)\to 0$.  This is because, in the Nambu notation, the lower right
element  of Eq.~(\ref{Green's function}) can be regarded as a Green's function with a line possessing
opposite momentum from the upper left element, i.e., its momentum is in the opposite direction.  This means that, 
in addition to ladder diagrams, we must include maximally-crossed diagrams shown in Fig.~\ref{fig6}(d).

The conclusion of the preceding remarks is that the set of diagrams needed to minimally generalize the usual $\sf{T}$-matrix approximation
to a system with pairing among the spins-$\uparrow$ is given by Fig.~\ref{fig6}(b). The first term contains the ladder
series and the second term contains the crossed diagrams Fig.~\ref{fig6}(d).  This gives:
\begin{equation}
\label{t and tc matrix self energy}
\boldsymbol{\Sigma}^{}_{\uparrow}(k)=\sum_{q}{\bf T}(k+q)G^{}_{\downarrow}(q)+\sum_{q}{\bf T}_{\rm c}(q-k)G^{}_{\downarrow}(q),
\end{equation}
where 
\bea
\label{matrix T-matrix}
{\bf T}(k)&=&\boldsymbol{\sigma}_{z}\big[\openone\lambda^{-1}+\boldsymbol{\Pi}_{A}(k)\big]^{-1},
\\
\label{mc t matrix}
{\bf T}_{\rm c}(k)&=&\lambda^{2}\boldsymbol{\sigma}_{z}\boldsymbol{\Pi}^{2}_{B}(k)\big[\openone\lambda^{-1}+\boldsymbol{\Pi}_{B}(k)\big]^{-1}
\eea
Here, the inverse is understood as a matrix inverse and the bubbles $\boldsymbol{\Pi}_{A}(k)$  and $ \boldsymbol{\Pi}_{B}(k)$ are  given by 
\bea
\label{Pi A}
\boldsymbol{\Pi}_{A}(k)&=&\sum_{q}{\bf G}_{\uparrow}(q)G^{}_{\downarrow}(k-q)\boldsymbol{\sigma}_{z},
\\
\label{Pi B}
\boldsymbol{\Pi}_{B}(k)&=&\sum_{q}{\bf G}_{\uparrow}(q)G^{}_{\downarrow}(k+q)\boldsymbol{\sigma}_{z}. 
\eea
Having derived the natural generalization of the ladder approximation to incorporate the possibility of intraspecies
pairing among the spins-$\uparrow$, we now show that, in the unitary regime, it cannot possess any off-diagonal
pairing correlations but merely reproduces the polaron Fermi liquid, implying that we must go beyond this 
level of approximation.  To do this, we make the replacement Eq.~(\ref{lambda vs a}) and take the limit
 $\Lambda \to \infty$ while holding $\as$ fixed (a procedure that, above, we argued to be valid away from
the weak-coupling BCS limit).  The Nambu  $\sf{T}$-matrix is thus 
\begin{equation}
\label{matrix T-matrix with a}
{\bf T}(k)=\boldsymbol{\sigma}_{z}\bigg[\openone \frac{4\pi \as}{m}+\boldsymbol{\Pi}_{A}(k)-\openone\sum_{\bf k}^{\Lambda}\frac{1}{2\epsilon_{\bf k}}\bigg]^{-1},
\end{equation}
where, when we take the limit $\Lambda \to \infty$, it is clear that the last term yields a divergence proportional to the identity matrix.  To be nonzero, 
a similar divergence must appear in $\boldsymbol{\Pi}_A(k)$; however, because of the Nambu structure of ${\bf G}_\uparrow(q)$, only the upper left component of $\boldsymbol{\Pi}_{A}$
can possess such a divergence.  Thus, all elements of ${\bf T}(k)$ except the upper-left will vanish, and we obtain: 
\begin{equation}
\label{unitarity matrix t matrix}
\lim_{\Lambda\rightarrow \infty}{\bf T}(k)=\left(\begin{array}{cc}{\sf T}(k) & 0 \\0 & 0\end{array}\right),
\end{equation}
where ${\sf T}(k)$ is the usual  $\sf{T}$-matrix, satisfying 
\bea
\label{unitarity t matrix}
&&\big[{\sf T}({\bf k},i\omega_{n})\big]^{-1}=\frac{4\pi \as}{m} 
\\
&&+\frac{1}{\beta}\sum_{{\bf q},i\nu_{n}}G_{\uparrow}({\bf q},i\nu_{n})G_{\downarrow}({\bf k}-{\bf q},i\omega_{n}-i\nu_{n})-\sum_{\bf q}\frac{1}{2\epsilon_{\bf q}}.
\nonumber 
\eea
To be clear, the quantity $G_{\uparrow}({\bf q},i\nu_{n})$ appearing in this formula is the upper-left component of the full matrix Green's function ${\bf G}_\uparrow(\bq,i\nu_n)$.
A similar simplification occurs in ${\bf T}_{\rm c}(k)$, which, when we make the same replacement and take the 
limit of $\Lambda \to \infty$, yields
\begin{equation}
\lim_{\Lambda\rightarrow \infty}{\bf T}_{\rm c}(k)=\left(\begin{array}{cc}0 & 0 \\0 & -{\sf T}(-k)\end{array}\right),
\end{equation}
giving for the self-energy 
\be
\label{Eq:selffl}
\boldsymbol{\Sigma}_{\uparrow}(k)=\sum_{q}\left(\begin{array}{cc}{\sf T}(q+k) & 0 \\0 & -{\sf T}(q-k)\end{array}\right)G^{}_{\downarrow}(q),
\ee
which, crucially is {\it diagonal\/} in Nambu space and, thus, possesses no pairing correlations.  In fact, Eq.~(\ref{Eq:selffl}) is
exactly the self-energy within the usual \lq\lq polaron\rq\rq\ Fermi liquid description of strongly imbalanced Fermi gases, equivalent
to the Chevy variational wavefunction~\cite{ChevyPRA06} as shown in Ref.~\onlinecite{CombescotPRL07}.

  Thus, although we generalized the ladder approximation in the simplest possible way to include 
pairing correlations (summing ladder and crossed-ladder
diagrams but with a {\it Nambu\/} spin-$\uparrow$ Green's function), we have found that, within this approximation, 
such pairing correlations are not stable.  The results of this section, however, tell us two things: Firstly, we must consider sets of diagrams
that go beyond this approximation.  Secondly, to have a result that is finite when we make the replacement Eq.~(\ref{lambda vs a}) and take $\Lambda
\to \infty$, we must consider diagrams with ladder or crossed-ladder type subdiagrams.

\subsection{Combined ladder and crossed-ladder diagrams}

\begin{figure}
\includegraphics[width= \columnwidth]{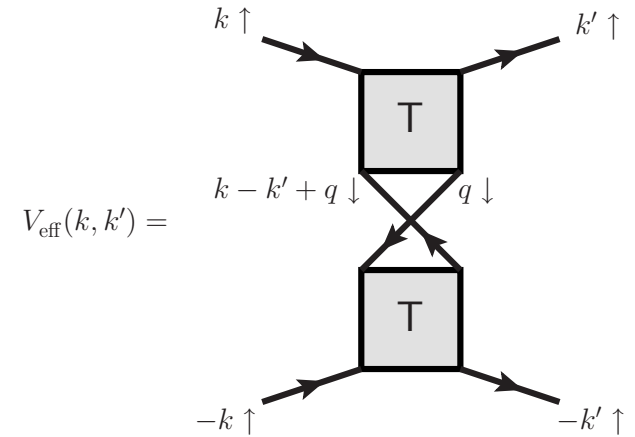}%
\caption{The effective induced interaction between spins-$\uparrow$, including contributions to all orders 
in perturbation theory, that is the many-body generalization of the leading order effective interaction 
shown in Fig.~\ref{fig3} }
\label{fig7}
\end{figure}

As we have seen, the ladder and crossed-ladder (or maximally-crossed) sets of diagrams contain geometric sums that are nonzero when we exchange
the bare coupling $\lambda$ for the scattering length $\as$ using Eq.~(\ref{lambda vs a}) and take the limit $\Lambda\to \infty$.  Since 
the quantities ${\bf T}(k)$ and ${\bf T}_{\rm c}(k)$ are diagonal in this limit, each possessing only one nonzero element, the corresponding
contributions to the self-energy in Fig.~\ref{fig6}(b) are also diagonal.  However, there exist additional sets of diagrams containing
${\bf T}(k)$  and  ${\bf T}_{\rm c}(k)$ as subdiagrams that are {\it not\/} diagonal in this limit; the simplest such self-energy
diagrams are shown as the final two diagrams in Fig.~\ref{fig6}(c).  Our inclusion of these diagrams may alternatively be understood
on physical grounds as due to the fact that paired superfludidity mixes particles and holes: roughly speaking,  ${\bf T}$ describes
particle-particle scattering and  ${\bf T}_{\rm c}$ describes hole-hole scattering.  Including pairing requires mixing these, suggesting
the incorporation of these diagrams.  
%
%
Thus, Fig.~\ref{fig6}(c) represents the full set of self-energy diagrams
that we consider here: The ladder and crossed-ladder terms that yield the polaron Fermi liquid (as shown above) and the combined ladder and
crossed-ladder diagrams that, as we now show, capture the instability of the polaron Fermi liquid to $p$-wave pairing.  
This self energy is:
\begin{widetext}
\begin{align}
\label{full self-energy}
\boldsymbol{\Sigma}^{}_{\uparrow}(k)&=\sum_{q}{\bf T}(k+q)G^{}_{\downarrow}(q)+\sum_{q}{\bf T}_{\rm c}(q-k)G^{}_{\downarrow}(q)\nonumber\\&-\lambda^{2}\boldsymbol{\sigma}_{z}\sum_{q,q'}\boldsymbol{\Pi}_{A}(k+q)\big[\openone\lambda^{-1}+\boldsymbol{\Pi}_{A}(k+q)\big]^{-1}{\bf G}_{\uparrow}(q)\boldsymbol{\sigma}_{z}\boldsymbol{\Pi}_{B}(q'-q)\big[\openone\lambda^{-1}+\boldsymbol{\Pi}_{B}(q'-q)\big]^{-1}G_{\downarrow}(q'-q+k)G_{\downarrow}(q')\nonumber\\&-\lambda^{2}\boldsymbol{\sigma}_{z}\sum_{q,q'}\boldsymbol{\Pi}_{B}(q'-k)\big[\openone\lambda^{-1}+\boldsymbol{\Pi}_{A}(q'-k)\big]^{-1}{\bf G}_{\uparrow}(q)\boldsymbol{\sigma}_{z}\boldsymbol{\Pi}_{A}(q+q')\big[\openone\lambda^{-1}+\boldsymbol{\Pi}_{A}(q+q')\big]^{-1}G_{\downarrow}(q+q'-k)G_{\downarrow}(q').
\end{align}
\end{widetext}
Again exchanging $\lambda$ for $\as$ using Eq.~(\ref{lambda vs a}) and    letting $\Lambda\rightarrow\infty$, we get 
\begin{equation}
\lambda\boldsymbol{\sigma}_{z}\boldsymbol{\Pi}_{A}(k)\big[\openone\lambda^{-1}+\boldsymbol{\Pi}_{A}(k)\big]^{-1}\rightarrow \left(\begin{array}{cc}{\sf T}(k) & 0 \\0 & 0\end{array}\right)
\end{equation}
and 
\begin{equation}
\lambda\boldsymbol{\sigma}_{z}\boldsymbol{\Pi}_{B}(k)\big[\openone\lambda^{-1}+\boldsymbol{\Pi}_{B}(k)\big]^{-1}\rightarrow \left(\begin{array}{cc}0 & 0 \\0 & -{\sf T}(k)\end{array}\right),
\end{equation}
so that Eq.~\eqref{full self-energy} reduces to
\begin{equation}
\label{self energy near unitarity}
\boldsymbol{\Sigma}_{\uparrow}(k)=\sum_{q}\left(\begin{array}{cc}{\sf T}(q+k)G^{}_{\downarrow}(q)
 & V^{}_{\rm eff}(k,q)F^{}_{\uparrow}(q) \\V^{}_{\rm eff}(k,q)F^{*}_{\uparrow}(q) & -{\sf T}(q-k)G^{}_{\downarrow}(q)
\end{array}\right),
\end{equation}
where the effective pairing interaction $V^{}_{\rm eff}(k,k')$ is shown diagrammatically in Fig.~\ref{fig7} and is explicitly given by
\begin{equation}
\label{effective interaction at unitarity}
V^{}_{\rm eff}(k,k')=\sum_{q}{\sf T}(q+k){\sf T}(q-k')G_{\downarrow}(k-k'+q)G_{\downarrow}(q). 
\end{equation}
The effective interaction 
Eq.~\eqref{effective interaction at unitarity} is a
direct generalization of the induced interaction of the weak BCS limit
\eqref{perturbative interaction}, except now with a energy and
momentum dependent coupling, i.e., the bare interaction $\lambda$ in
Eq.~\eqref{perturbative interaction}
 has
been replaced by the $\sf T$-matrix in Eq.~\eqref{effective interaction at
unitarity}.

With this effective interaction, the gap equation is 
\be
\label{Eq:generalgap}
\Delta_\uparrow(k) = \sum_q V_{\rm eff}(k,q) F_\uparrow(q),
\ee
where, as in the weak-coupling case, in principle $\Delta_\uparrow(k)$ is frequency-dependent, 
although we shall again assume it to be static.  
Unlike the weak-coupling regime, however, now the effective interaction also depends on 
the pairing gap via the relationship, Eq.~(\ref{unitarity t matrix}), of the ${\sf T}$ matrix to the spin-$\uparrow$ 
Green's function.  This difficult self-consistency problem simplifies near the assumed
continuous phase transition, where $\Delta_\uparrow(\bk)$ vanishes, a regime we now focus on.  
In this regime,  we can simply approximate the ${\sf T}$ matrices by their
form in the imbalanced normal phase.


In principle all of the Green's functions appearing in the effective
interaction Eq.~\eqref{effective interaction at unitarity} are the exact
interacting Green's functions.  To proceed, we make some simplifying 
approximations, firstly by  replacing these with their 
noninteracting expressions (equivalent to assuming the imbalanced
Fermi liquid phase is noninteracting) and 
 neglecting the external 
frequency dependence of Eq.~\eqref{effective interaction at unitarity} 
(i.e., setting $\omega_k = \omega_k' \to 0$).  This gives
\bea
&&V_{\rm eff}(\bk,\bk') = 2T \sum_\Omega \int \frac{d^3q}{(2\pi)^3}
{\sf T}(\bq+\bk,i\Omega) {\sf T}(\bq-\bk',i\Omega) 
\nonumber \\
&&\qquad\qquad\qquad \times 
\frac{1}{i\Omega - \xi_{\bk-\bk'+\bq\downarrow}}
\frac{1}{i\Omega -\xi_{q\downarrow} }.
\label{Eq:veff2}
\eea
%
The sum over Matsubara frequencies can be evaluated using the standard trick~\cite{Mahan}
that requires  the location of the poles of the summand in the complex plane.
This is simplified by our knowledge that poles of the normal-state $\sf T$-matrix on the real axis occur at the onset of
$s$-wave paired superfluidity.  Since we are studying the strongly imbalanced limit,
we can assume no such poles contribute, 
 and proceed by  keeping only the poles from the
Green's functions in the second line of Eq.~(\ref{Eq:veff2}).  This leads to
\bea
&&V_{\rm eff}(\bk,\bk') = 2T \sum_\Omega \int \frac{d^3q}{(2\pi)^3}
{\sf T}(\bq+\bk,\xi_{\bq\downarrow}) {\sf T}(\bq-\bk',\xi_{\bq\downarrow}) 
\nonumber \\
&&\qquad\qquad\qquad \times 
\frac{\nf(\xi_{\bq\downarrow})}{ \epsilon_q  - \epsilon_{\bk-\bk'+\bq}},\label{Eq:veff3}
\eea
our final result for the static induced interaction for the majority spins-$\uparrow$, 
mediated by the spins-$\downarrow$, in a strongly interacting imbalanced Fermi gas. 

Our main results come from numerically evaluating Eq.~(\ref{Eq:veff3}).  However, it is useful to 
first make a bit more approximate analytic progress by invoking an approximation that is valid in 
the large-imbalance limit.  Thus,  to evaluate the remaining integral over $\bq$, we begin by noting
that the Fermi function restricts, at low $T$, the momentum to $q<k_{{\rm F}\downarrow}$,
inside the spin-$\downarrow$ Fermi surface.  But since we're interested in the
regime where $\bk$ and $\bk'$ are on the spin-$\uparrow$ Fermi surface, 
in the strongly imbalanced limit $k_{{\rm F}\uparrow} \gg k_{{\rm F}\downarrow}$ the
momentum argument of the ${\sf T}$-matrices in Eq.~(\ref{Eq:veff3})  is approximately given simply by
$k_{{\rm F}\uparrow}$.  In this strongly imbalanced limit the first ${\sf T}$-matrix in 
Eq.~(\ref{Eq:veff3}) simplifies to 
\be
  {\sf T}(\bk_{\rm F},\xi_{\bq\downarrow})\simeq \Big(\frac{m}{4\pi \as}-\frac{m k^{}_{{\rm
F}\uparrow}}{4\pi^{2}}\Big)^{-1},
\ee
with a similar expression holding for the second ${\sf T}$-matrix in 
Eq.~(\ref{Eq:veff3}).  Within this approximation, valid at large population 
imbalance $P\to 1$, the ${\sf T}$-matrices are thus independent of momenta.
As can be seen in the original expression Eq.~(\ref{self energy near unitarity}),
when this occurs $V_{\rm eff}$ is simply proportional to the bare bubble occuring in
 the weakly interacting regime (see Eq.~(\ref{perturbative interaction})).  Evaluating
the remaining integrals and making the projection to the $p$-wave channel
(as we did in the weakly interacting regime), we obtain
\begin{equation}
\label{asymptotic Tc}
k^{}_{\rm B}T_{\rm c}\approx \frac{2{\rm e}^{\gamma}}{\pi}\epsilon^{}_{{\rm F}\downarrow}
\exp\Big[-\frac{3}{2}\frac{z}{\ln(z)}\Big(\frac{\pi}{2k^{}_{{\rm F}\downarrow}\as}-\frac{z}{2}\Big)^{2}\Big],
\end{equation} 
with $z= k^{}_{{\rm F}\uparrow}/k^{}_{{\rm F}\downarrow}$, a result that we emphasize is only
valid in the asymptotic strongly imbalanced regime. 

\begin{figure}
 \vspace{-0.5cm}
\includegraphics[width= \columnwidth]{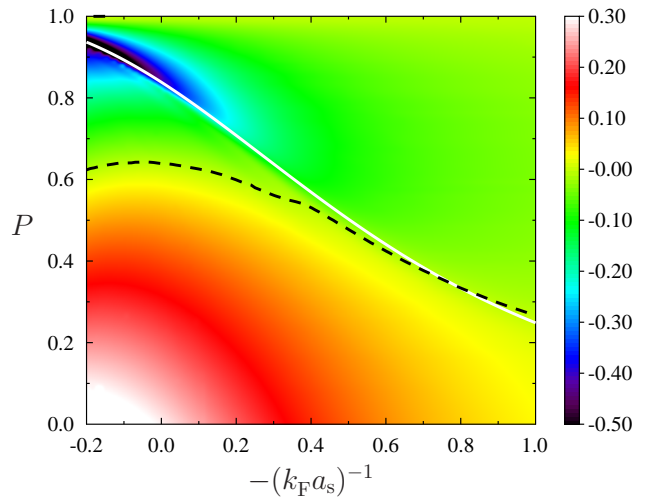}%
\caption{ (Color online)  The $p$-wave channel of the effective
interaction between majority spins, $v^{1}_{k^{}_{\rm F\uparrow},k^{}_{\rm F\uparrow}}$,
multiplied by the Fermi-energy density of states $N_{\uparrow}({\epsilon}^{}_{{\rm
F}\uparrow})$, as a function of the density
imbalance $P$ and $s$-wave scattering length $\as$, is shown.
Here, $k^{}_{\rm F}=(k^{}_{{\rm F}\uparrow}+k^{}_{{\rm
F}\downarrow})/2$. At large $P$, above the dashed line,
it is {\it attractive\/} leading to a $p$-wave superfluid at
temperatures below $T^{}_{\rm c}$.  The solid white line labels 
the location of a line of FFLO quantum critical points, defined by ${\sf T}({\bf Q}_{\rm FFLO},0)\rightarrow\infty$, at zero temperature.  This coincides  with the location where 
$v^{1}_{k^{}_{\rm F\uparrow},k^{}_{\rm F\uparrow}}$ is large.\label{fig10}}
\end{figure}

 Away from this $P\to 1$ limit, we must 
perform a numerical evaluation of the integral in Eq.~(\ref{Eq:veff3}) and the $p$-wave
projection; this yields a dimensionless effective interaction 
$
{v^{1}_{k_{{\rm F}\uparrow},k_{{\rm F}\uparrow}} N_{\uparrow}(\epsilon_{{\rm F}\uparrow})}$ that we 
plot in Fig.~\ref{fig10} and in the top panel of Fig.~\ref{fig9}.  For this calculation
we assumed the $T\to 0$ limit; since our resulting transition temperature is still small compared to
the spin-$\uparrow$ Fermi energy this should be an accurate approximation.  
These figures show that we find attractive  interactions in the $p$-wave channel over a wide
range of the interactions and the population imbalance.

\begin{figure}
\vspace{-0.5cm}
\includegraphics[width= \columnwidth]{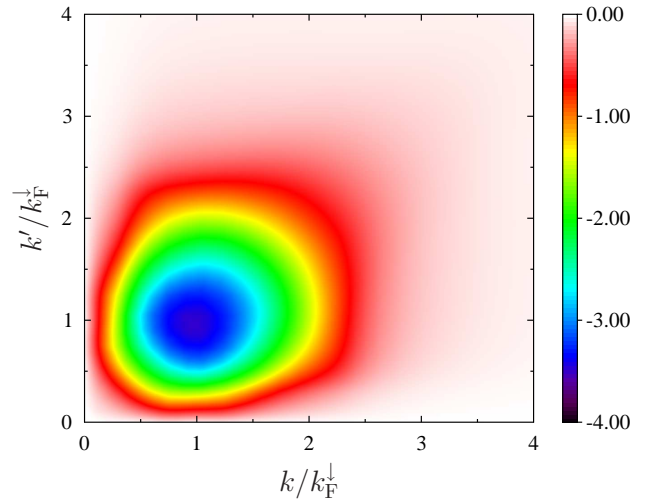}%
\caption{ 
(Color online) 
The momentum dependence of
the $p$-wave channel $v^{1}_{k,k'}$ of the effective interaction,
Eq.~\eqref{effective interaction at unitarity}, as defined by
Eq.~\eqref{angular momentum channels}, for zero frequency and at
unitarity  for a polarization of $P=0.85$. As one can see the
interaction is nonzero only within a window of approximately $k=k'\pm
k_{{\rm F}\downarrow}$.  
} \label{fig8}
\end{figure}

\begin{figure}
\vspace{0.5cm}
\includegraphics[width= \columnwidth]{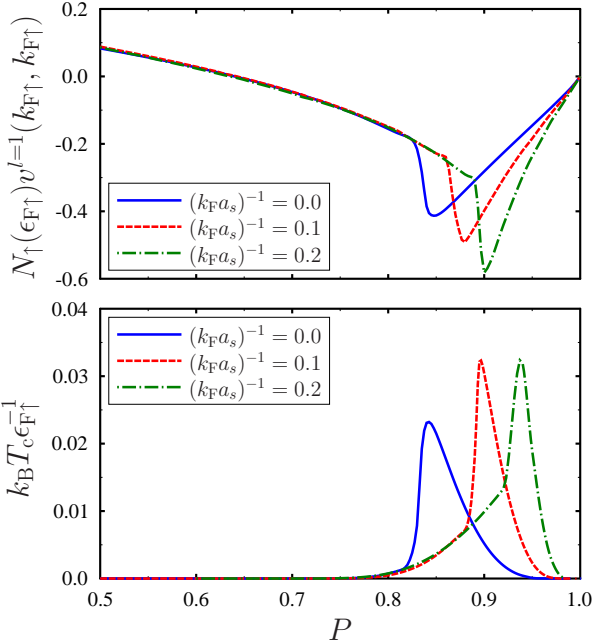}%
\caption{(Color online) The top panel shows the on-shell $\ell= 1$ channel of the effective interaction \eqref{effective interaction at unitarity} for spin-$\uparrow$ fermions, times the density of states, as a function of polarization, at unitary and into the BEC side. The bottom panel shows the corresponding $p$-wave transition temperature,
according to Eq.~(\ref{transition temperature}), with  $\epsilon_{\rm F\uparrow}/\epsilon_{\rm F\downarrow} = [(1+P)/(1-P)]^{2/3}$. \label{fig9}}
\end{figure}

  The corresponding transition temperature, plotted in the bottom panel of Fig.~\ref{fig9}
as well as in the phase diagram Fig.~\ref{phasefig}, is given by the same formula
as in the weak-coupling limit, i.e., 
\begin{equation}
\label{transition temperature2}
k_{\rm B}T_{{\rm c}}\approx 
\frac{2\epsilon_{{\rm F}\downarrow}e^{\gamma}}{\pi}\exp\Big[\frac{1}{v^{1}_{k_{{\rm F}\uparrow},k_{{\rm F}\uparrow}} 
N_{\uparrow}(\epsilon_{{\rm F}\uparrow})}\Big],
\end{equation}
where to arrive at this formula we used the fact that, as in the weak-coupling limit, $v^{1}_{k,k'}$ 
(the $p$-wave projection of Eq.~(\ref{Eq:veff3})) is only nonzero for $k$ and $k'$ close to each other, 
within  a window of approximately  $\pm\epsilon^{}_{{\rm F}\downarrow}$.  
We have verified this numerically; a plot displaying the typical behavior of 
$v^{1}_{k,k'}$  is shown in Fig.~\ref{fig8}. 

\begin{figure}
\includegraphics[width= \columnwidth]{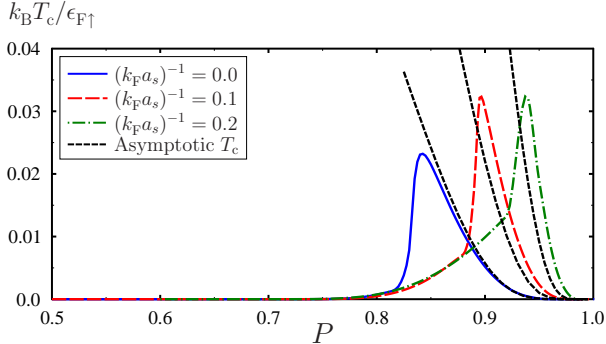}%
\caption{(Color online) A comparison of the transition temperature
obtained from Eq.~\eqref{asymptotic Tc}, valid for asymptotically
large polarizations, and a full numerical treatment of the induced
interaction Eq.~\eqref{effective interaction at unitarity}, using
Eq.~(\ref{transition temperature2}), with  $\epsilon_{\rm
F\uparrow}/\epsilon_{\rm F\downarrow} =
[(1+P)/(1-P)]^{2/3}$. \label{fig11}}
\end{figure}

It is interesting to compare the value of $T_{\rm c}$ obtained via a direct numerical evaluation of 
Eq.~(\ref{Eq:veff3}) to our approximate asymptotic formula Eq.~(\ref{asymptotic Tc}).  In 
Fig.~\ref{fig11} we plot these as a function of population imbalance for three different values of the
scattering length.  These curves show that, while Eq.~(\ref{asymptotic Tc}) is accurate asymptotically
close to $P = 1$, it misses the peak and drop in $T_{\rm c}$ that generically occurs with 
decreasing $P$.

What is the origin of this peak in the predicted $T_{\rm c}$ occuring for very large imbalance?  
Certainly, one expects an increase in $T_{\rm c}$ as $P$ decreases from unity, arising from the
 increase in the density of  spins-$\downarrow$ which provide the induced interactions.  
More precisely, the effective interactions among the spins-$\uparrow$ 
are mediated by particle-hole excitations of the spins-$\downarrow$.  However, particle-hole excitations
with a wavevector larger than the diameter of the spin-down Fermi surface ($2k_{{\rm F}\downarrow}$) are energetically 
suppressed, implying that the density response function is strongly varying for $k\simeq 2k_{{\rm F}\downarrow}$.  
This strong variation as a function of momentum leads to a large $p$-wave projection of the induced
interaction when $k\simeq k_{{\rm F}\downarrow}$ (since, for $p$-wave pairing, we require an attractive interaction
that strongly varies around the spin-$\uparrow$ Fermi surface).  If we then assume that the maximum $T_{\rm c}$ will
occur when $k_{{\rm F}\uparrow} = 2k_{{\rm F}\downarrow}$, we are led to the prediction that $T_{\rm c}$ will peak near
\be
\label{Eq:predictedpeak}
P = \frac{k_{{\rm F}\uparrow}^3 - k_{{\rm F}\downarrow}^3}{ k_{{\rm F}\uparrow}^3 + k_{{\rm F}\downarrow}^3} = \frac{7}{9}\simeq 0.78,
\ee
close to the value of the peak position shown in Fig.~\ref{fig9}.

We note that Eq.~(\ref{Eq:predictedpeak}) only {\em approximately\/} locates the peak position; indeed, Fig.~\ref{fig9} shows some variation
of the peak position as a function of interactions.  (We have not pushed this computation deep into the BEC regime
where we know the magnetic superfluid ground state intervenes~\cite{SR2006,SR2007}.)
Indeed, the preceding argument is strictly true in the
weak-coupling limit where the effective interaction is given by the Lindhard function $L(x)$ that has a singularity 
near $x =1$ that leads to a similar peak in $T_{{\rm c}}$ in the weak coupling limit. (Although, as mentioned above,
the weak-coupling $T_{{\rm c}}$ is orders of magnitude smaller than the values plotted here.)
 In the strong-coupling 
limit, we expect this argument to still approximately hold since there the Luttinger theorem ensures that the presence of spin-$\uparrow$
and spin-$\downarrow$ Fermi surfaces at the same volume as in the weakly interacting limit (until the broken symmetry phase 
appears)~\cite{Luttinger,SachdevYang}. 

We also note an additional possible reason for the occurence of a peak in $T_{\rm c}$ at large imbalance: A proximate FFLO 
instability.  Thus, one expects an instability towards FFLO pairing for imbalanced Fermi gases, occuring when
there is a divergence of the retarded $\sf T$ matrix at zero frequency (but nonzero wavevector $\bQ_{\rm FFLO}$).
In the vicinity of such a phase transition, the $\sf T$ matrix would, correspondingly, possess a large magnitude
that would enhance the effective induced interactions.  To test this, in Fig.~\ref{fig10}, we plot, as a white line, the $P$ at
which such a {\it zero temperature\/} FFLO instability would first occur; as seen in this plot it closely correlates to 
the regime where the $p$-wave $T_{\rm c}$ is largest.  Our calcultion of this FFLO transition of course neglected
the possibility of $p$-wave pairing among the spins-$\uparrow$.  Thus, if the $p$-wave transition occurs first,
with decreasing temperature, it would likely move the location of the FFLO phase boundary.  Further detailed analysis
will be required to sort out these various competing instabilities.

We conclude this section by briefly justifying the use of the 
 on-shell approximation, in which we assume that we only need
the $p$-wave induced interactions for momenta on the spin-$\uparrow$ 
Fermi surface.  As is known from the theory of superconductivity, such an approximation is
typically only valid for weakly interacting quasiparticles and,   near unitarity,
 particles of opposite spin are strongly interacting.  Despite this, in
highly imbalanced systems the lifetime of the quasiparticles remains
extremely long; this implies the quasiparticle-quasiparticle
interaction is still quite weak.   This can be seen directly from the
experimental and theoretical  results of Ref.~\cite{SchirotzekPRL09}.
In Ref.~\cite{SchirotzekPRL09} the Fermi liquid properties of systems
with an imbalance as low as $P\approx0.70$ were explained well within
a theoretical framework consisting of only a single spin-$\downarrow$
quasiparticle, which indicates induced interactions of like spins, is
subdominant to such things as the renormalization of the chemical
potential and effective mass, at least for $T\gtrsim T_{\rm c}$.  
Thus, we argue that our analysis is valid if the strongly imbalanced
nonsuperfluid phase is truly a Fermi liquid for $T\gtrsim  T_{\rm c}$.

\section{Low-temperature pairing gap}
\label{sec:ltpg}
In the present section our goal is to estimate the magnitude of the $p$-wave
pairing gap at $T\to 0$.  One subtlety is that 
our method for calculating the induced attraction among the spins-$\uparrow$ 
is only valid for $T>T_{c}$, since we neglect the pairing among the 
spins-$\uparrow$ in computing the $\sf T$ matrix.  However, if we assume
that the induced pairing $\Delta_\uparrow(\bk)$ is small in magnitude compared
to $\epsilon_{\rm F\uparrow}$ (as one might expect), then we can assume 
$\Delta_\uparrow(\bk)$ has only a small effect on the $\sf T$ matrix, and 
proceed to neglect it when estimating $\Delta_\uparrow(\bk)$.  

The relevant equation for the low-temperature gap is Eq.~(\ref{on-shell gap equation}),
but with $V_{\rm eff}(\bk,\bq)$ given by Eq.~(\ref{Eq:veff3}).  Focusing on the 
$p$-wave channel, we have, for $T\to 0$,
\be
\label{eq:lowtgap}
\Delta_{\uparrow} (\bk) =  - 3\sum_{\bq} v^{1}_{k,q}\kh\cdot \qh \frac{\Delta_{\uparrow}(\bq)}{2E_\bq}.
\ee
Following the work of Anderson and Morel, we expect the dominant $p$-wave instability to be 
of the $p_x+ip_y$ form~\cite{AndersonMorelPR61}.  We thus write
\be
\label{fullgap}
\Delta_{\uparrow}(\bk) = \Delta(k^2) (\kh_x +i\kh_y).
\ee
Converting the sum in Eq.~(\ref{eq:lowtgap}) to an integral (recall we set the system volume to unity),
 we obtain
\be
\Delta(k^2) 
 =  - \frac{3}{16 \pi^2} \int_0^\infty q^2 dq\, \Delta(q^2) v^{1}_{k,q}\frac{1}{|\xi_{q\uparrow}|}F\big[\frac{|\xi_{q\uparrow}|}{\Delta(q^2)} \big],
\label{eq:deltakxky4}
\ee
where the function 
\be
F[x]\equiv x^2\Big[1+x\big( \frac{1}{x^2} -1\big) \tan^{-1}\frac{1}{x}\Big],
\label{eq:f1}
\ee
arises from evaluating the angular integration.
We now use the fact that $v^1_{k,q}$ is sharply peaked near $k=q$, and assume the rest of the integrand of Eq.~(\ref{eq:deltakxky4}) is
smooth there.  This yields the approximate formula 
\bea
1
&\simeq &- \frac{3}{16 \pi^2}\bar{v}_k  \frac{k^2}{|\xi_{k\uparrow}|}F_{1}\big[\frac{|\xi_{k\uparrow}|}{\Delta(k^2)} \big],
\label{eq:deltakxky5}
\\
\label{eq:veek}
\bar{v}_k &\equiv &\int_0^\infty dq v^{1}_{k,q} .
\eea
 Restricting attention to the vicinity of the Fermi surface by setting $k\to k_{\rm F \uparrow}$, and using the asymptotic 
value of $F[x] \to \frac{\pi}{2} x$ for $x\to 0$, we have 
\be
\Delta(k_{\rm F\uparrow}^2) \simeq - \frac{3}{32 \pi} \bar{v}_{k_{\rm F \uparrow}} k_{\rm F \uparrow}^2.
\ee
We can further simplify this formula by noting that, since $v^{1}_{k,q}$ is peaked for $k=q$ with a 
width of approximately $\pm k_{\rm F\downarrow}$, we have 
$\bar{v}_{k_{\rm F \uparrow}} \simeq 2 k_{\rm F\downarrow}  v^{1}_{k_{\rm F \uparrow},k_{\rm F \uparrow}}$.  Then, 
normalizing $\Delta(k_{\rm F}^2)$ to the Fermi energy and using Eq.~(\ref{nsigma}), 
we have 
\be
\frac{\Delta(k_{\rm F}^2)}{\epsilon_{\rm F\uparrow}} \simeq - \frac{3\pi}{4} \frac{k_{\rm F\downarrow}}{k_{\rm F\uparrow}} 
N_\uparrow(\epsilon_{{\rm F}\uparrow}) v^1(k_{\rm F\uparrow},k_{\rm F\uparrow}).
\label{Eq:maxgap}
\ee
The product of the last two factors of 
Eq.~(\ref{Eq:maxgap}) is precisely
what is plotted in Fig.~\ref{fig9}.  At unitarity, $\as^{-1} = 0$, this
factor reaches $\simeq 0.4$ at $P \simeq 0.85$, or $\frac{k_{\rm F\downarrow}}{k_{\rm F\uparrow}} \simeq 0.43$.
Plugging these values into Eq.~(\ref{Eq:maxgap}) gives the estimate $\frac{\Delta(k_{\rm F}^2)}{\epsilon_{\rm F\uparrow}} \approx 0.4$.
Although this estimate is rather large, it is important to keep in mind that this is the maximum pairing gap; the full gap function 
will exhibit nodes according to Eq.~(\ref{fullgap}).  An important issue for future work is to find a more accurate estimate for
the low-$T$ pairing gap for the majority species of an imbalanced Fermi gas.

\section{Experimental detection}
\label{sec:ed}
We argue that, although small (a few percent of the Fermi energy),
the transition temperature is within range of 
 current experimental
capabilities \cite{ShinPRL08}. However, the experimental
detection of such a state may still be challenging. 
One striking way to identify the presence of superfluidity is via the 
presence of vortices in a rapidly-rotating cloud, as done in 
Ref.~\onlinecite{ZwierleinNature05} to detect $s$-wave pairing correlations.
However, in Ref.~\onlinecite{ZwierleinNature05} the detection of BCS pairs was accomplished by ramping
the magnetic field onto the BEC side of the resonance, changing them into
molecular pairs.  For the case of the $p$-wave Cooper pairs predicted here, it is 
not clear that such a ramp is possible.

Owing to this difficulty, here we focus on two other possible ways
to detect  $p_x+ip_y$ Cooper pairing among the majority species of
an imbalanced Fermi gas: Radio-Frequency (RF) spectroscopy (in which 
pairing correlations are detected by a shift in the rate at which atoms in a 
particular state are transferred to a third unoccupied level~\cite{Chin2004}) and
noise correlations (studied theoretically
in Ref.~\onlinecite{AltmanPRA04} and experimentally in 
Ref.~\onlinecite{Greiner2005}).

\subsection{RF spectroscopy of the majority spin $p$-wave superfluid}
\begin{figure}
\vspace{0.5cm}
\includegraphics[width= \columnwidth]{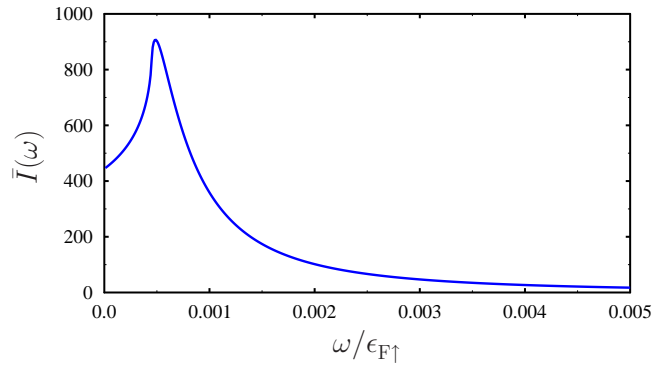}
\caption{(Color online) The normalized zero temperature  RF line-shape \eqref{RF
spectrum},
$\bar{I}(\omega)=I_{\uparrow}(\omega)/\int_{0}^{\infty}d\omega\,
I_{\uparrow}(\omega)$, of a $p$-wave superfluid, with a $p_{x}+ip_{y}$
ground state.  The
magnitude of the gap is approximated by the transition temperature,
$\Delta_{0}\approx 0.03 \epsilon^{}_{{\rm F}\uparrow}$ and the
chemical potential $\mu_{\uparrow}\approx\epsilon^{}_{{\rm F}\uparrow}
$, for $P\approx 0.85$ at unitarity.  In general for a $p_{x}+ip_{y}$
ground state the peak occurs near
$\omega\simeq\sqrt{\mu^{2}_{\uparrow}+\Delta^{2}_{0}}-\mu_{\uparrow}$,
and as $\omega\rightarrow\infty$,
$\bar{I}(\omega)\rightarrow\omega^{-3/2}$, while for
$\omega\rightarrow 0$, $\bar{I}(\omega)\rightarrow
(\omega+2\mu_{\uparrow})^{5/2} $. \label{fig12}}
\end{figure}

 To use RF spectroscopy to probe $p_x+ip_y$ pairing of the spins-$\uparrow$,  
one would measure the number of spin-$\uparrow$ atoms transferred 
 to a known unoccupied level, as a function of
applied RF frequency \cite{ChenRepProgPhys09}.  Neglecting final state
effects, the total transferred  at a given frequency is given by  
\begin{equation}
\label{RF spectrum} I_{\sigma}(\omega)\propto \sum_{\bf
k}A_{\sigma}({\bf k},\xi^{}_{\bf k\sigma}-\omega)n^{}_{\rm
F}(\xi^{}_{\bf k\sigma}-\omega),
\end{equation}
where the spectral function $A_{\sigma}(k)$ is related to the imaginary part of the retarded Green's function $G^{\rm r}_{\sigma}(k)$ by
\begin{equation}
A_{\sigma}(k)=-\frac{1}{\pi}{\rm Im}\, G^{\rm r}_{\sigma}(k),
\end{equation}
and $n^{}_{\rm F}(\omega)$ is the Fermi distribution function. 
It is convenient to write  the superfluid Green's function in the mean-field spectral representation:
\begin{equation}
\label{mean-field p-wave Green's function}
G^{\rm r}_{\uparrow}({\bf k},\omega)=\frac{u^{2}_{\bf k}}{\omega-E_{\bf k}+i\eta}+\frac{v^{2}_{\bf k}}{\omega+E_{\bf k}+i\eta},
\end{equation}
where the so-called coherence factors are
\bse 
\begin{align}
\label{u and v}
u^{2}_{\bf k}&=\frac{1}{2}\bigg(1+\frac{\xi_{\bf k\uparrow}}{E_{\bf k}}\bigg),\nonumber\\
v^{2}_{\bf k}&=\frac{1}{2}\bigg(1-\frac{\xi_{\bf k\uparrow}}{E_{\bf k}}\bigg),
\end{align}
\ese
and $E_{\bf k}=\sqrt{\xi^{2}_{\bf k\uparrow}+|\Delta_\uparrow({\bf k})|^{2}}$.   The spectral function is then
\begin{equation}
\label{mean field superfluid spectral function}
A_{\uparrow}({\bf k},\omega)=u^{2}_{\bf k}\delta(\omega-E_{\bf k})+v^{2}_{\bf k}\delta(\omega+E_{\bf k}).  
\end{equation}
As we have discussed, the $p$-wave ground state is expected to have $p_x+ip_y$ symmetry, given by Eq.~(\ref{fullgap})\cite{AndersonMorelPR61,GurariePRL05,NishidaAnalPhys09}.  
A full calculation of the RF lineshape requires $\Delta(k^2)$ for all $k$, a difficult self-consistency problem that is beyond the scope
of this work.  We proceed by simply assuming $\Delta(k^2)=\Delta_0$, i.e., a constant value.  
%
%
%
  At zero temperature and using Eqs.~\eqref{u and v}, \eqref{mean field superfluid spectral function}, 
and \eqref{fullgap}
 the momentum integrals in Eq.~\eqref{RF spectrum} can be done exactly, although the result is too unwieldy to present here. 
 Figure~\ref{fig12} shows the resulting RF line shape at unitarity.
In this plot we chose $\Delta_0 \simeq \kb T_{\rm c}$, assuming the magnitude of the pairing gap reflects the transition
temperature (with the latter given by $T_{\rm c}\simeq .03\epsilon_{\rm F\uparrow}/\kb$, a typical maximum value in Fig.~\ref{fig9}).
Such an estimate is more conservative than the rather large estimate found in Sec.~\ref{sec:ltpg}.
As can be seen from Fig.~\ref{fig12}, unlike the $s$-wave state, which has a hard gap for $\omega\lesssim \Delta^{2}_{0}/\epsilon_{\rm F\uparrow}$, the RF line shape of this $p$-wave state remains non-zero for all $\omega$, 
due to the nodes in the pairing gap.   In principle one could use this to detect the $p$-wave phase of imbalanced Fermi gases.  One issue, however, is the small magnitude of the peak position in the energy in
Fig.~\ref{fig12}, which occurs at a scale of order $\Delta_0^2/\epsilon_{\rm F\uparrow}$ (rather than at $\Delta_0$).  To see whether $p$-wave pairing is truly observable, we need a good estimate of the low-temperature 
pairing gap in this phase. 

\subsection{Density correlations}

The use of  spatial correlations in the density of the free expanded gas, or density-density correlations, as a tool to probe the order of the superfluid state of ultracold atomic systems was put forth in Ref.~\onlinecite{AltmanPRA04}.  As free expansion of the density probes the momentum distribution of the trapped system, spatial correlations of this expansion probes correlations in momentum space.  It is these momentum correlations of the superfluid state that are a direct consequence of the Cooper pairing. 

Theoretically the quantity of interest is the equal time density-density correlation function of the spins-$\uparrow$
\begin{equation}
D({\bf r},{\bf r}',t)={\rm Tr}\rho^{}_{\rm trap}\, \delta\hat{n}_{\uparrow}({\bf r},t)\delta\hat{n}_{\uparrow}({\bf r}',t),
\end{equation}
where $\delta\hat{n}_{\uparrow}({\bf r})=\hat{n}_{\uparrow}({\bf r})-\langle\hat{n}_{\uparrow}({\bf r})\rangle$ and $\rho^{}_{\rm trap}$ is the equilibrium density matrix of the trapped system, i.e., superfluid, while the time evolution of the operators is given by free un-trapped Hamiltonian. As the spins-$\downarrow$ are assumed unpaired, their momentum correlations are essentially featureless and won't be included.     Neglecting the inhomogeneity caused by the trapping, we take the density matrix to be given by the zero temperature BCS state,
\begin{equation}
|\Psi_{\rm BCS}\rangle=\prod_{\bf k}\left(u^{}_{\bf k}+v_{\bf k}a^{\dagger}_{{\bf k}\uparrow}a^{\dagger}_{-{\bf k}\uparrow}\right)|{\rm vac}\rangle,
\end{equation}
with the coherence factors given by Eq.~(\ref{u and v}) above.
 Within this approximation and assuming the density is
measured on space and times scales much larger than $k^{-1}_{{\rm
F}\uparrow}$ and $\epsilon^{}_{{\rm F}\uparrow}$, we find 
\begin{align} D({\bf r},{\bf r}',t)&\approx\left(\frac{m}{2\pi
t}\right)^{6}\sum_{{\bf q},{\bf q}'}\phi^{*}_{\bf q}({\bf
r}m/t)\phi^{}_{\bf q}({\bf r}'m/t)\nonumber\\&\times
\phi^{}_{{\bf q}'}({\bf r}'m/t)\phi^{*}_{{\bf q}'}({\bf r}m/t)|u^{}_{\bf q}|^{2}|v^{}_{{\bf q}}|^{2},
\end{align} 
where $\phi^{}_{{\bf q}}({\bf k})$ is the Fourier transform of the single-particle wave function of the trapped system with quantum index ${\bf q}$.  Typically these will be very sharply peaked near ${\bf k}$ and ${\bf k}'$, thus
\begin{equation}
\label{approximate den-den correlation}
D({\bf r},{\bf r}',t)\approx\left(\frac{m}{2\pi t}\right)^{6}\tilde{\delta}({\bf r}m/t+{\bf r}'m/t)|u^{}_{{\bf k}({\bf r})}|^{2}|v^{}_{{\bf k}({\bf r})}|^{2},
\end{equation} where $\tilde{\delta}({\bf r})$ is essentially a
broadened delta function,  the form of which depends on the specific
details of the trapped system.   Excluding  the pre-factor, the weight
of the ``delta function''  $|u^{}_{{\bf k}({\bf r})}|^{2}|v^{}_{{\bf
k}({\bf r})}|^{2}$ provides the information about the momentum
correlations with ${\bf k}({\bf r})={\bf r}m/t$. 

\begin{figure}
\includegraphics[width= 0.99\columnwidth]{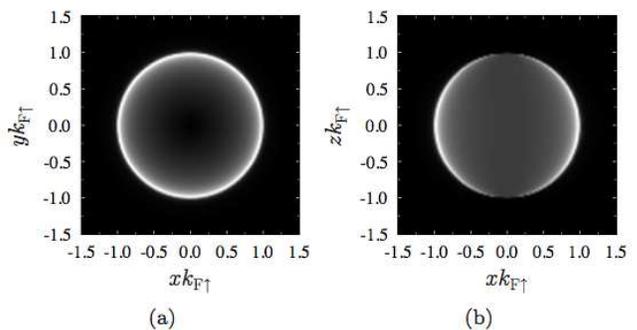}             
 \caption{The column integrated density-density correlations \eqref{column integrated density-density}  (weight of delta function only) of a free expanded $p^{}_{x}+ip^{}_{y}$ superfluid with maximum gap $ \Delta_{0}= 0.03 \epsilon^{}_{{\rm F}\uparrow}$ and symmetry axis $\hat{\bf z}$ is shown: $(a)$ by integrating along the $z$-axis and $(b)$ integrating along the $y$-axis.  As the product of $|u^{}_{{\bf k}({\bf r})}|^{2}|v^{}_{{\bf k}({\bf r})}|^{2}$ appearing in \eqref{column integrated density-density} is only appreciably nonzero near the Fermi surface, the magnitude of the correlations is maximum there. This results in the ring-shape seen in both (a) and (b).   }
 \label{fig14}
\end{figure}

Experimentally one does not measure the local density in 3-space, but
instead the column integrated density. For example if the detector in
located in the $x$-$y$ plane at a distance $z_{0}$ from the origin
then the column integrated density-density correlation function is  
\begin{align}
\label{column integrated density-density}
&D({\bf r}^{}_{\bot},{\bf r}'_{\bot},t)=\int\limits_{-\infty}^{z_{0}}dz dz'\,D({\bf r},{\bf r}',t)\nonumber\\&\approx \left(\frac{m}{2\pi t}\right)^{6}\tilde{\delta}({\bf r}_{\bot}m/t+{\bf r}'_{\bot}m/t)\int\limits_{-\infty}^{z_{0}}dz\,|u^{}_{{\bf k}({\bf r})}|^{2}|v^{}_{{\bf k}({\bf r})}|^{2},
\end{align}
where ${\bf r}_{\bot}$ corresponds to spatial directions perpendicular to the column integration.
 As the $p$-wave breaks rotational symmetry, here chosen to be in the $\hat{\bf z}$ direction, the observed column integrated density-density correlations depends on the relative orientation of the spontaneously broken symmetry direction and the detector.  Figure \ref{fig14} shows the  column integrated density-density correlation (weight function only) for a detector located in an $x$-$y$ plane and an $x$-$z$ plane for the $p^{}_{x}+ip^{}_{y}$ state with a gap $ \Delta_{0}= 0.03 \epsilon^{}_{{\rm F}\uparrow}$. In Fig.~\ref{fig14}b the $p$-wave nature can clearly be seen when the integrated column density is obtained by integrating along the direction that is perpendicular to the symmetry of the order parameter.  

\section{Conclusions}
\label{sec:conc} 

Cold atom experiments have demonstrated the capability to study a remarkably
simple many-body physics problem: That of two species of attractively
interacting fermion as a function of the interatomic scattering length and
relative densities of the two species.  Despite the simplicity of this problem,
the resulting phase diagram is quite rich, showing regions of phase separation,
imbalanced superfluid, and normal Fermi liquid.

The question we pursue here is, why aren't there more phases of imbalanced
Fermi gases?  Indeed, experiments on imbalanced Fermi gases observe the absence of any
broken symmetry phases for a large range of parameters.  For example, 
at unitarity, Ref.~\cite{ShinNature} finds that phase separation
vanishes above $P\simeq 0.36$.  Are there truly no broken-symmetry ground states of
imbalanced Fermi gases over the range of polarization values 
$0.36\alt P<1$, or do other phases lurk at low temperatures in this
strongly interacting system?

This paper partially addresses such questions by proposing that, in the large imbalance
region the true ground state is a $p$-wave superfluid of the spin-$\uparrow$ fermions, with
the order setting in below a temperature, $T_{\rm c}$, plotted in 
Fig.~\ref{phasefig}.  Important questions for futher work include obtaining more accurate estimates
for $T_{\rm c}$ and studying the $p$-wave gap equation at low $T$, and determining the transition
temperature for the onset
of pairing for the spins-$\downarrow$ in the unitariy regime (and finding how this onset modifies
the properties of the spin-$\uparrow$ Cooper pairs).  An additional question concerns finding
more experimental signatures of the onset of $p$-wave pairing, to help ascertain the validity of this scenario.

 From a general point of view, 
a natural question is whether other phases, such as higher-angular
momentum superfluids or FFLO phases intervene in the $T\to 0$ limit of imbalanced Fermi gases.  Since
the $p$-wave interaction becomes repulsive for small imbalance (below the dashed line in Fig.~\ref{fig10}),
it is likely that other angular momentum channels can become dominant in this regime.  Additionally,
although the window of FFLO stability is extremely thin for the simplest FFLO-type state within
mean-field theory, it is possible that generalized FFLO states, within theoretical approaches
that go beyond mean-field theory, can lead to a wider regime of FFLO stability~\cite{Yoshida,LeoAshvin,LeoPRA}.

\begin{acknowledgments}
KRP would like to thank Hartmut Hafferman and Herbert Fotso for useful discussions. 
This work was supported by the Louisiana Board of Regents, under grant LEQSF (2008-11)-RD-A-10.
  
\end{acknowledgments}

\end{document}